\documentstyle[floats,aps,psfig,12pt]{revtex}
\begin{document}
\title{$J/\Psi$-Production at Photon-Photon Colliders as a Probe of the
Color Octet Mechanism}
\par\vskip20pt
\author{J.P. Ma \and B.H.J. McKellar \and C.B. Paranavitane } 
\date{\today}
\draft
\preprint{UM-P-97/46 RCHEP97/07}
\address{{\it Research Center for High Energy Physics, School of Physics,
University of Melbourne, Parkville, Victoria 3052,
     Australia}}
\vskip 30pt

\begin{center}
\maketitle{Abstract} 
\end{center}
\begin{abstract}
We study $J/\Psi$ production at photon-photon colliders, which can be realised
with Compton scaterring of laser photons at $e^+e^-$ colliders. 
We find that the production rate through the color-octet channel is 
comparable to that through the color-singlet channel. 
Experimentally the two mechanisms can be studied separately 
because the processes have different signals.  
\end{abstract}
\newpage

It is commonly believed that predictions for quarkonium production 
can be made with perturbative QCD at a  certain level because 
the mass of the heavy  quarks provide a large scale. 
Before the work of Bodwin et al \cite{B1}, such predictions were based on 
perturbative calculations
for the production of a heavy quark pair and on some assumptions 
of the formation of a quarkonium from this pair. This formation 
was described by models such as the color singlet model and the color 
evaporation model.  
Recently, based on non relativistic QCD(NRQCD), 
a factorized form for quarkonium production rates 
has been proposed\cite{BFL}, where the formation 
is described systematically by a series of NRQCD matrix elements.  
Such a formalism is possible because the heavy quark, $Q$, in the rest frame
of the  quarkonium 
moves with a small velocity $v$, hence an expansion 
in this small velocity is possible. Predictions based 
on this factorized form take not only the effects of a heavy quark pair, which 
have the same quantum numbers as those of the quarkonium, 
into account, but also those effects from other possible 
components of the quarkonium 
with different quantum numbers, such as a heavy quark pair in a
color-octet state. Although effects of color-octet states 
for $J/\Psi $ production 
are at higher orders in the small velocity expansion than that of
a color-singlet state, they can be significant phenomenologically 
and can  even 
overwhelm that of the  color-singlet states in certain 
kinematic regions. Actually, after including the effect 
of color-octet states, the Tevatron data on prompt charmonium 
production can  be fitted well to theoretical predictions\cite{BY}. 
However, it is difficult to identify the detailed significance 
of color-octet states in a process where one or two hadrons 
are involved in the initial state. Besides possible 
inaccuracies introduced by finite orders of perturbation theory, 
there are many uncertainties in theoretical predictions related 
to the initial hadrons     
such as parton distributions and higher twist effects\cite{Ma}, etc. 
Hence it is important to study the effect of color-octet
states at non-hadronic colliders, where theoretical predictions 
are more certain than those for  hadronic colliders. 
Such  studies are carried out for $e^+e^-$-colliders in \cite{EE}. 
\par
In this work we study $J/\Psi$ production at photon-photon 
colliders. Usually, photon-photon collisions always 
happen at e.g. $e^+e^-$ colliders, and the photon beams 
can be described in the Weiz\"acker-Williams approximation. 
But the luminosity of such beams is greatly reduced 
from that of the original collider. It is pointed out  
in \cite{GKT} that one can use Compton scattering of laser light 
to obtain $\gamma\gamma$ collisions at an $e^+e^-$ collider
with approximately the same luminosity as that of the $e^+e^-$ beams.
Unlike those  
in Weiz\"acker-Williams approximation the photons obtained through 
Compton(back) scattering of laser light are hard, i.e., they 
have a large probability of carrying a large fraction of the beam 
energies. With these photon beams 
the total cross-section due to $\gamma\gamma$ collisions 
can be calculated by: 
\begin{equation}
\sigma_{\rm tot} = \int dx_{1} dx_{2} f_\gamma (x_{1}) f_\gamma (x_{2})
        \sigma_{\gamma\gamma} (x_1,x_2),
\end{equation}
where 
$\sigma_{\gamma\gamma} (x_1,x_2)$ is the cross-section due to 
photon-photon collisions in which one photon has  a fraction 
$x_1$, another has $x_2$ of the initial beam energy. $f_\gamma (x)$ is the 
photon distribution, 
given by:
\begin{eqnarray}
f_\gamma (x) &=&  \frac{1}{N} \{1-x + \frac{1}{1-x} - \frac{4 x}{x_m (1-x)} +
\frac{4 x^{2}}{x_m^{2} (1-x)^{2}}\}, \\ 
\nonumber
 N &=&  (1-\frac{4}{x_{m}}-\frac{8}{x_{m}^{2}}) \ln (1+x_{m})
+ \frac{1}{2} + \frac{8}{x_{m}}- \frac{1}{2 (1+x_{m})^{2}} , 
\end{eqnarray}
with
\begin{equation}
x_{m} = \frac{4E \omega_{o}}{m_{e}^{2}} {\rm cos} ^{2} \frac{\theta} {2}, 
\end{equation} 
where $E$ is the beam energy, $\omega_{o}$ is the energy of the laser 
photon and  $\theta$ is the angle between the directions of the beam and the 
laser photon. The energy fraction $x$, carried by a scattered photon is 
restricted 
in the range:
\begin{equation}
   0 \le x  \le {\frac{x_m}{1+x_m}}. 
\end{equation}
We will consider the photon beams realised at the proposed 
NLC\cite{NLC} and take $\omega_0=1.26{\rm eV}$ and $\alpha =0$ for 
our numerical results. 
\par
With two photons as an initial state the charmonium $J/\Psi$ can be produced 
via the following processes at the leading order of coupling constants 
$\alpha$ and $\alpha_s$: 
\begin{eqnarray} 
   \gamma & + & \gamma \rightarrow J/\Psi +G, \\  
    \gamma &  + & \gamma \rightarrow J/\Psi +\gamma. 
\end{eqnarray}
In the first process a $^3 S_1\  c\bar c$-pair in a color-octet state is 
produced and then the pair is transformed  into $J/\Psi$, while 
in the second, a $^3 S_1\  c\bar c$-pair in a color-singlet state 
is produced and then transformed into $J/\Psi$. The transition 
in the first process happens at order of $v^4$ and in the second 
at order of $v^0$ in the small velocity expansion. Although 
the probability for the transition from 
a color-octet $c\bar c $ is smaller than that from a color-singlet 
$c\bar c$, the first process is enhanced by $\alpha_s$ relative 
to $\alpha$ and it has a cross-section comparable to the second 
processes. It should be mentioned that at order $v^4$ there are 
other possible color-octet states which can be transformed into 
$J/\Psi$, but these states can not be produced at the order of coupling 
constants which we consider. 
Further, these two processes can have different experimental signatures.
The quarkonium in the final state is accompanied by a 
jet initiated 
by the gluon in the first process and by an  energetic photon 
in the second. If these processes are observed, 
one may easily identify whether the $J/\Psi$ comes from a color-octet 
$c\bar c$ pair or not.  Hence studying 
these processes can help us to understand more about 
the color-octet mechanism. 

The effect of color-octet states in 
quarkonium production at  hadron colliders is studied extensively  
\cite{CL},\cite{KLS}. Unlike at hadron colliders where 
many subprocesses  
can contribute to the quarkonium production we have here only one subprocess. 
It is straight forward to calculate the cross-sections of the two processes. 
One of the Feynman diagrams for them is given in Fig.1. 
For the first process we have: 
\begin{eqnarray}
\label{eq:shortcross}
\frac{d{\sigma_{\gamma\gamma}  
}}{d\hat t}  & = &  {\frac{16 M_{J/\Psi}}{3}} ({\frac {2}{3}})^4 (4\pi)^2  
\alpha_{s} \alpha^{2}  \langle 0 | O_8^{J/\Psi} (^3S_1) |0\rangle
 \nonumber \\ 
  & &  \cdot \frac{\hat s^{2}(\hat s-M_{J/\Psi}^{2})^{2} 
 + \hat t^{2}(\hat t-M_{J/\Psi}^{2})^{2}+ \hat u^{2}(\hat u-M_{J/\Psi}^{2})^{2}
} {\hat s^{2} (\hat s-M_{J/\Psi }^{2})^{2} (\hat t-M_{J/\Psi}^{2})^{2} 
(\hat u-M_{J/\Psi }^{2})^{2}}.
\end{eqnarray}
The variables $\hat s$, $\hat t$ and $\hat u$ are standard Mandelstam 
variables, $M_{J/\Psi}$ is twice the $c$ quark mass, $m_c$. 
The matrix element $\langle 0 | O_8^{J/\Psi} (^3S_1) |0\rangle$
is defined in NRQCD in \cite{BFL}  and represents 
the probability of  transition from a 
color-octet $^3S_1$ $c\bar c$ into the $J/\Psi$. The differential cross-section 
for the second process can be obtained by the replacement in Eq.(7) via
\begin{equation} 
  \langle 0 | O_8^{J/\Psi} (^3S_1) |0\rangle \rightarrow 
  {\frac {4\alpha}{9 \alpha_s}}  \langle 0 | O_1^{J/\Psi} (^3S_1) |0\rangle,  
\end{equation} 
where the matrix element $ \langle 0 | O_1^{J/\Psi} (^3S_1) |0\rangle$ 
represents the effect of the transition of a color-singlet $c\bar c$ pair, 
its definition can be found in \cite{BFL} also.
\par
We take two possible values for the center of mass energy, $\sqrt s$, 
at the proposed NLC, $\sqrt s=500{\rm GeV}$ and $1000 {\rm GeV}$.
We obtain the following results for the total cross-section for the process 
in Eq.(5):
\begin{eqnarray}
\sigma_{\rm tot} (J/\Psi + G)  &=& 2.66{\frac { \langle 0 | O_8^{J/\Psi} 
(^3S_1) |0\rangle}
      {\rm (GeV)^3 } } {\rm (pb)}, \ \ \ \  {\rm for} \ 
     \sqrt s=500{\rm GeV} \\    
\sigma_{\rm tot} (J/\Psi + G)  &=& 0.46 {\frac { \langle 0 | O_8^{J/\Psi} (^3S_1) |0\rangle}
      {\rm (GeV)^3 } } {\rm (pb)}, \ \ \ \  {\rm for} \ 
     \sqrt s=1{\rm TeV}. 
\end{eqnarray}
In the above results we have taken $\alpha=1/128$, $\alpha_s (m_c)=0.3$ 
and $m_c=1.5(\rm GeV)$. 
The final value of the cross-section
depends on the matrix element. 
This matrix element is extracted from different processes, but its value varies 
from $0.006({\rm GeV)^3}$ to $0.02({\rm GeV})^3$   
from the different processes reflecting uncertainties in theoretical 
predictions. If we take $0.01({\rm GeV})^3$ for the matrix element, the 
cross-section 
is $0.0266{\rm pb}$ at $\sqrt s=500{\rm GeV}$ and $0.0046{\rm pb}$ at $\sqrt s=1{\rm TeV}$.
With a proposed luminosity for the NLC of 
$100{\rm fb}^{-1}$ per year at $\sqrt s=500{\rm GeV}$,
there will be several hundred events which can be detected through 
the leptonic decay of $J/\Psi$. Although the event number is not large, 
once these
events are observed, there is  direct access to the matrix element.

The cross-section 
through the color-singlet channel  can be estimated with the replacement in Eq.(8). The color-singlet
matrix element can be approximately  extracted from the leptonic decay of $J/\Psi$. We take 
its value as $1.1{\rm (GeV)}^3$ and obtain the cross-section: 
\begin{eqnarray}
\sigma_{\rm tot} (J/\Psi + \gamma)  &=& 0.031  
      {\rm (pb)}, \ \ \ \  {\rm for} \
     \sqrt s=500{\rm GeV} \\
\sigma_{\rm tot} (J/\Psi + \gamma )  &=& 0.005 
      {\rm (pb)}, \ \ \ \  {\rm for} \
     \sqrt s=1{\rm TeV}.
\end{eqnarray}
With these estimates one can conclude that 
the $J/\Psi$ production through the color-octet channel will be
comparable to  that through the color-singlet channel. 
The production through these mechanisms can be distinguished 
experimentally. 
\par
Having studied the total cross-section we turn to some distributions for 
the process in Eq.(5). The distributions studied here are the same 
for the process in Eq.(6). In Fig.2 we give the differential cross-section
$\frac {d \sigma_{\rm tot}} {d p_{t}^{2}}$ as function of
$p_{t}^{2}$, where $p_{t}$ is the 
transverse momentum of the $J/\Psi$. 
We define a total cross-section with the $J/\Psi$ having a $p_{t}$ 
larger than $p_{\rm min}$ as:
\begin{equation} 
   \sigma_{\rm tot} (p_{\rm min} ) =\int _{p_{\rm min}}^{p_{\rm max}} dp_t 
    \frac {d \sigma_{\rm tot}} { d p_t}. 
\end{equation} 
$\sigma_{\rm tot} (p_{\rm min} )$ as function of $p_{\rm min}$ 
is drawn in Fig.3. 
From Fig.2 and Fig.3 one can see that most of $J/\Psi$ will be produced
with relatively  small $p_{t}$. The cross section $\sigma_{\rm tot}(p_{\rm min})$ 
decreases rapidly with increasing $p_{\rm min}$. 
\par   
For $J/\Psi$ production at hadronic colliders the process of gluon fragmentation 
plays an important role. The reason is 
because the contribution from that process is kinematically
much less suppressed than that from non-fragmentation processes. 
However in our case a gluon can be only produced via the diagrams 
given in Fig.1, hence it can be expected that the contribution 
from gluon fragmentation will have roughly the same suppression 
from the kinematics as that for the process in Eq.(5), and the gluon
fragmentation begins at the order of $\alpha_s$. Therefore we can 
expect that the contribution from gluon fragmentation will not be 
significant. 
Besides gluon fragmentation, quark fragmentation 
can also lead to contributions to inclusive $J/\Psi$ production, 
where the fragmentation is through color-octet mechanisms 
at the order of $\alpha_s^2$ and is 
studied in \cite{YM}. In our case the production 
of a quark is less suppressed kinematically than the process 
in Eq.(6), but the probability for quark fragmentation is very small. 
The ratio of the probabilities of quark and gluon fragmentation 
from  \cite{YM} is: 
\begin{equation} 
\frac {P(q\rightarrow J/\Psi)}{P(G\rightarrow J/\Psi)} 
   \approx 0.22 \alpha_s(2M_c).  
\end{equation}   
It may be expected that quark fragmentation will not lead
to dominant contributions to the inclusive production of $J/\Psi$.
However a more detailed study may be needed, which is beyond the scope
of this work. 
\par 
Our results for $J/\Psi$ production through the process in Eq.(5) 
can also be used for $\chi_{cJ}$, for $J=0,1,2$, by the replacement:
\begin{equation} 
  \langle 0 | O_8^{J/\Psi} (^3S_1) |0\rangle
   \rightarrow  \langle 0 | O_8^{\chi_{cJ}} (^3S_1) |0\rangle.
\end{equation}     
The matrix element $\langle 0 | O_8^{\chi_{c1}} (^3S_1) |0\rangle$
is approximately  $0.0073 GeV^3$\cite{BF}. 
With this estimate, the cross section for $\chi_{c1}$ production at the NLC at 
$\sqrt{s} = 500 {\rm GeV}$ is $0.078 {\rm pb}$. 
the production of this exited state
can be important for overall $J/\Psi$ production 
because $\chi_{c1}$ decays via $\chi_{c1} \rightarrow J/\Psi + \gamma$ 
with a branching ratio of $27$\%.
Unlike $J/\Psi$ production, no color singlet contribution exists  
for $\chi_{cJ}$ which gives a significant contribution.

\par
To summarise, we studied in this work  $J/\Psi$ production 
at a photon-photon collider. 
It was found that in addition to the color-singlet contribution, a color 
octet $c\bar c$ pair makes a significant contribution to the production rate.
Additionally, it is possible to differentiate between the two mechanisms 
by tagging the photon produced in the color-singlet process.
Hence a photon-photon collider will provide an ideal 
place to study the color-octet mechanism for quarkonium production.  
 
\par

\noindent
{\bf Acknowledgement:}
\par      
This work is supported by The Australian Research Council and the Australian 
Postgraduate Award. 

\newpage
\centerline{\bf Figure Captions}
\par
\noindent
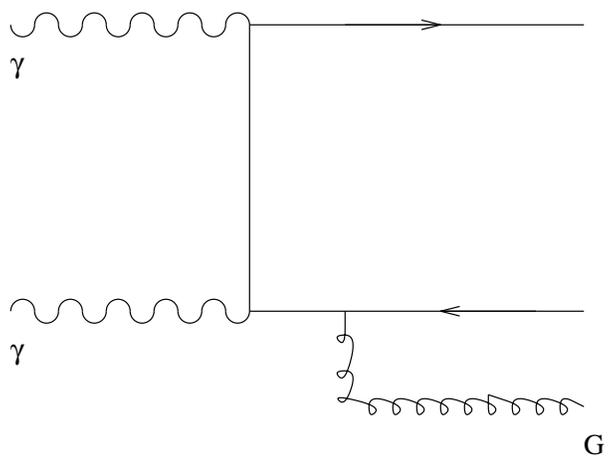
Fig.1: One of six Feynman diagrams for the production of  a       
color-octet $c\bar c$ pair. Others are obtained through permutations 
of photon and gluon lines. 
\par\noindent
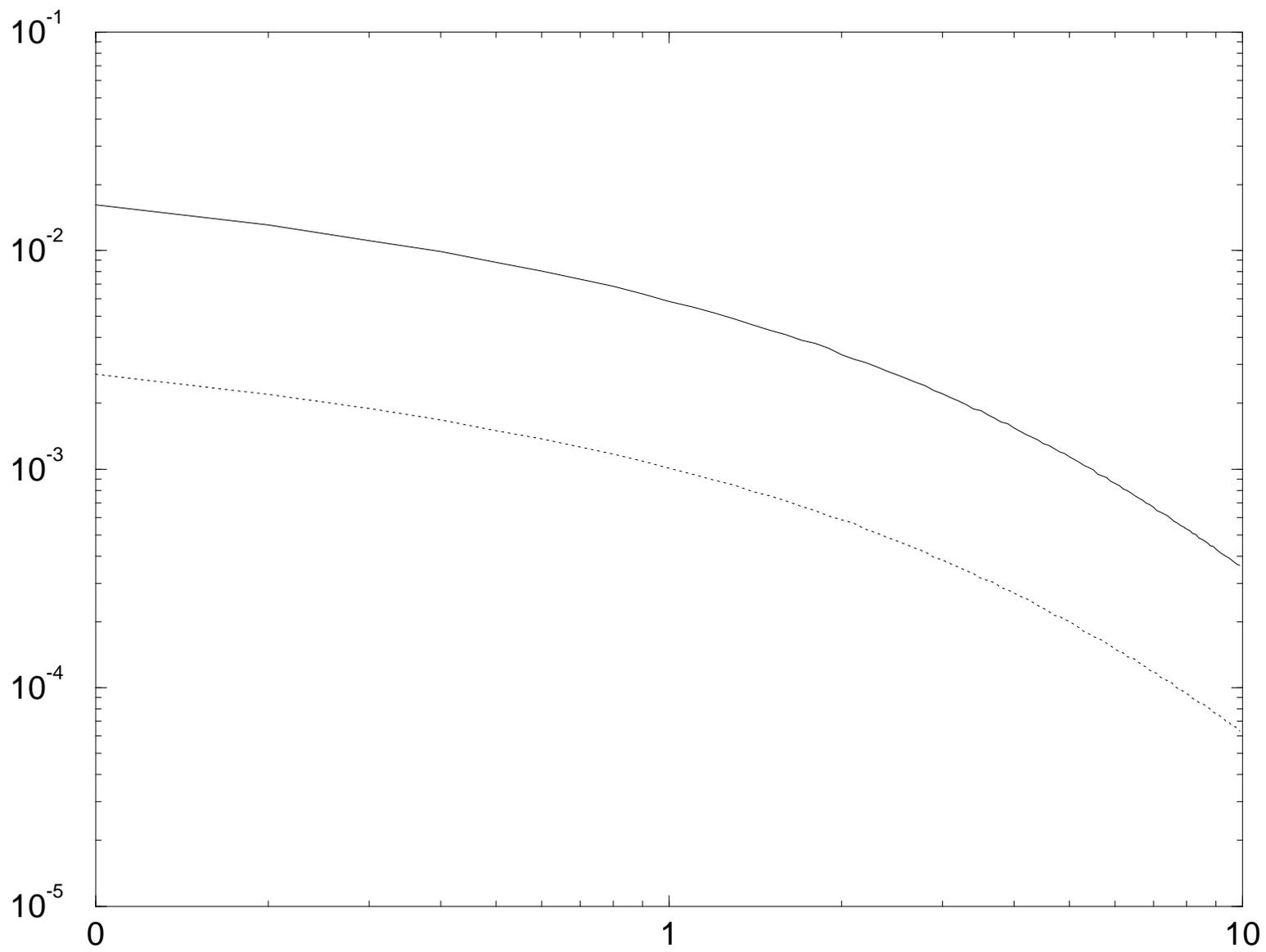
Fig.2: The color octet $p_{t}^{2}$-distributions as a function of 
$p_{t}^{2} ({\rm GeV}^{2})$ in ${\rm pb} {\rm GeV}^{-2}$.
The solid line is at $\sqrt{s} = 500 {\rm GeV}$ while the
dotted line is at $\sqrt{s} = 1{\rm TeV}$.
\par\noindent
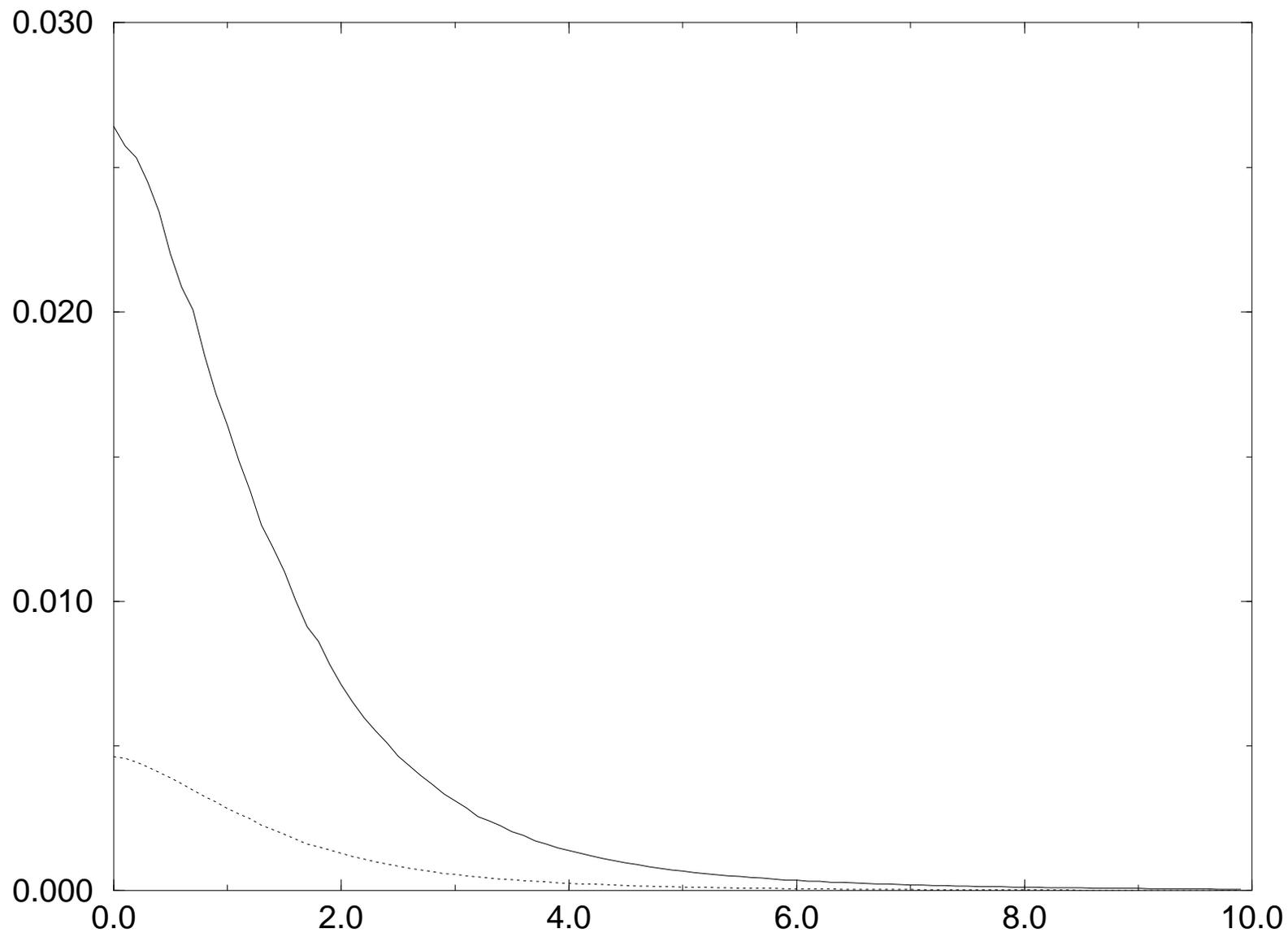
Fig.3: The color octet cross-section $\sigma(p_{\rm min}) ({\rm pb})$ 
as function of 
$p_{\rm min}({\rm GeV})$ 
as defined in Eq. (14). The solid and dotted lines are for  
$\sqrt{s} = 500{\rm GeV}$
and $\sqrt{s} = 1{\rm TeV}$ respectively.  
\par
\end{document}